\documentstyle[12pt]{article}
\begin{document}
\title{On the Ground State of Ferromagnetic Hamiltonians}
\author{J. Dittrich$^{a}$, V. I. Inozemtsev$^{b}$\\
{\small
$^{a}$ Nuclear Physics Institute, Academy of Sciences of the Czech
Republic}\\{\small
CZ-250 68  \v{R}e\v{z}, Czech Republic
\footnote{Also member of the Doppler
Institute of Mathematical Physics, Faculty of Nuclear Sciences and Physical
Engineering, Czech Technical University, Prague.}}
\\
{\small $^{b}$ BLTP JINR}\\
{\small 141980 Dubna, Moscow Region, Russia}}
\maketitle
\abstract{\sl It is generally believed that the ground state of the ferromagnetic
Heisenberg-Dirac-Van Vleck Hamiltonians acting on $s=\frac{1}{2}$ spins of a lattice with $N$ sites
has the maximal possible value of the total spin $S=\frac{N}{2}$ and is
$N+1$ times degenerate. We present a rigorous proof of this statement,
independent of the lattice dimension and topology.}
\newpage

\section{Introduction} Since the pioneering works on the
ferromagnetism by Heisenberg in 1928 \cite{Heisenberg}, Van Vleck
\cite{VanVleck} and others (see \cite{Mattisbook}, page 40 for a
brief history) much attention has been paid  to the spectrum of
the linear operators \begin{equation}
H=\frac{1}{2}\sum_{<i,j>}J_{ij}(\frac{1}{4}-{\bf s_i}\cdot{\bf
s_j}) \quad , \end{equation} where ${\bf s_j}=\frac{1}{2}{\bf
\sigma_j}$ are the spin-$\frac{1}{2}$ operators at the site
$j=1,\dots,N$ of a finite lattice (i.e. they are three-component
vectors of $2\times 2$-matrices) and $<i,j>$ indicates the
summation over all the pairs of nearest-neighbor lattice sites.
The properties of the ground state depend strongly on the sign of
coupling constants $\{ J_{ij}\}$. If all $J_{ij}<0$, the operator
is called antiferromagnetic. In 1962, Lieb, Mattis and Schultz
\cite{Lieb1, Lieb2} gave a rigorous proof that in
antiferromagnetic case the ground state $\psi_a$ has total spin
$S=0$ and is non-degenerate for even number of lattice sites. The
detailed structure of $\psi_a$ can be studied analytically only
for one-dimensional lattice with all $J_{ij}$ equal where the
model turns out to be solvable and its solution is given by the
Bethe Ansatz \cite{Bethe}. As for the higher dimensions, the
structure of $\psi_a$ can be analyzed by numerical methods only.

For the more transparent ferromagnetic case, all $J_{ij}>0$,
Mattis \cite{Mattis} considered as easy to show that the ground state
has the maximal spin $S=\frac{N}{2}$ and is $N+1$ times degenerate the
ground states being just state with all spins alligned up and its rotations.
However, as to our knowledge we did not see any fully detailed and
rigourous proof in the literature we are giving here the proof with
clearly formulated assumptions on the lattice of any dimension.

\section{The result and proof}
Before the main statement let us show a property of connected graphs
which will be used in its proof.
\\ \\ \noindent
{\bf Lemma.}
{\em Let $\Gamma$ be a connected graph with $N\geq 2$ vertices $1,2,\dots,N$ and
edges $<i,j>$ between some pairs of vertices $i,j=1,2,\dots,N$. Then there
exist two vertices $i_0\not= j_0$ such that the graphs
$\Gamma\setminus\{i_0\}$ and $\Gamma\setminus\{j_0\}$ obtained by removing
the vertices $i_0$ or $j_0$ respectively from $\Gamma$ (together with all
edges ending at them of course) are connected.}
\\ {\bf Proof.} Let us prove the Lemma by induction with respect to $N$. The
assertion is evidently valid for $N=2,3$. Let us now assume that it holds
for the graphs with numbers of vertices up to $N$ and that $\Gamma'$ is a
connected graph with $N+1$ vertices. Let $\Gamma''$ is obtained from
$\Gamma'$ by removing the vertex $1$.

If $\Gamma''$ is connected then
we can remove another two vertices $i_1,j_1$ leaving $\Gamma''$
connected by induction assumption. If both $<1,i_1>$, $<1,j_1>$ are edges of
$\Gamma'$ we can remove $1$ and e.g. $i_1$ leaving $\Gamma'$ connected.
If $<1,i_1>$ is while $<1,j_1>$ is not the edge of $\Gamma'$ we can remove
$1$ and $j_1$ leaving $\Gamma'$ connected. If non of $<1,i_1>$,
$<1,j_1>$ is the edge of $\Gamma'$ we can remove any of
$1,i_1,j_1$ leaving $\Gamma'$ connected.

If $\Gamma''$ has more than one connected components we can remove two
vertices from each one leaving it connected. From them at least one can
be chosen such that connection to vertex $1$ is not broken.  So we can
again remove at least two vertices from $\Gamma'$ leaving it connected.
\\ \hspace*{10.9cm}
\mbox{\rule[-1.pt]{6pt}{10pt}}
\\ \\
{\bf Theorem.}
{\em Let $\Gamma$ be a connected graph with $N\geq 2$ vertices,
\begin{eqnarray}
\label{Hdef}
H=\frac{1}{2}\sum_{<i,j>\subset\Gamma}
J_{ij}(\frac{1}{4}-{\bf s_i}\cdot{\bf s_j})
\\
J_{ij}>0 \quad\quad {\rm for}\; <i,j>\subset\Gamma
\end{eqnarray}
where the spin-$\frac{1}{2}$ operators at the site $i$ are given by
Pauli matrices as
${\bf s_i}=(\frac{1}{2}\sigma_{i,x},\frac{1}{2}\sigma_{i,y},
\frac{1}{2}\sigma_{i,z})$ for $i=1,\dots,N$.
Let $|\psi>$ be a ground state of the Hamiltonian $H$. Then
\begin{equation}
\label{Hpsi0}
H|\psi>=0
\end{equation}
and the following statements hold.
\\
a)
\begin{equation}
\label{Smax}
{\bf S}^2|\psi>=\frac{N}{2}\left(\frac{N}{2}+1\right)|\psi>
\end{equation}
with the total spin ${\bf S}=\sum_{i=1}^N{\bf s_i}$, i.e. $|\psi>$
is an eigenvector of ${\bf S}^2$ with the maximal eigenvalue.\\
b)
\begin{equation}
\label{sisj}
{\bf s_i\cdot s_j}\, |\psi>=\frac{1}{4}\, |\psi>
\end{equation}
for any pair $i\not= j$ where $i,j=1,\dots,N$.\\
c) There exist complex numbers $\alpha_k$ and matrices $u_k\in SU(2)$
($k=1,\dots,N+1$) such that
\begin{equation}
|\psi>=\sum_{k=1}^{N+1}\alpha_k\otimes_{i=1}^N u_k |\uparrow >
\end{equation}
where
\begin{equation}\nonumber
|\uparrow>=\left(\begin{array}{c} 1 \\ 0 \end{array}\right)
\end{equation}
is the spin-up eigenvector of the third component of one spin
(eigenvalue $\frac{1}{2}$). In other words, there are no other eigenvectors
$|\psi>$ satisfying (\ref{Hpsi0}) then those obtained from
$|\psi_0>=|\uparrow \dots \uparrow>=\otimes_{i=1}^N|\uparrow>$
(all spins up) by rotations and linear combinations.}
\\
{\bf Proof.}
As the possible value of the total momentum of two spins are $0$ and $1$,
the eigenvalues of $\frac{1}{4}-{\bf s_i\cdot s_j}$ are $0$ and $1$.
Therefore for any state $|\varphi>$ of $N$ spins
\begin{equation}
\label{twospins}
<\varphi|(\frac{1}{4}-{\bf s_i\cdot s_j})|\varphi>\,\geq\,0
\end{equation}
where the equality holds if and only if
$(\frac{1}{4}-{\bf s_i\cdot s_j})|\varphi>\,=\,0$.
So $0$ is the ground state energy of $H$ as we know that really
$H|\psi_0>=0$.

Let us further realize that statement b) implies a) by direct
calculation \begin{eqnarray} \nonumber {\bf S}^2|\psi>&=& \left[
\sum_{i=1}^N{\bf s_i}^2 + 2\sum_{i<j}{\bf s_i\cdot s_j}
\right]|\psi> \\ \nonumber &=&
\left[N\cdot\frac{3}{4}+2\cdot\frac{N(N-1)}{2}\cdot\frac{1}{4}\right]
|\psi> \\ &=& \frac{N}{2}\left(\frac{N}{2}+1\right) |\psi> \quad .
\end{eqnarray} The assertion c) follows from a) as the all states
satisfying (\ref{Smax}) can be obtained from $|\psi_0>$ by
rotations and linear combinations and the Hamiltonian is
rotationally invariant. To see this, realize that the group SU(2)
has an irreducible representation with given eigenvalue of the
total spin ${\bf S}^2$ and that there is just one state $|\psi_0>$
with the maximal third component of total spin and therefore just
one copy of this irreducible representation.

It remains to prove the assertion b) by induction with
respect to $N$. The assertion holds for $N=2$. Let us now assume
that the statement b) and consequently also a) holds for some integer
number $N\geq 2$. Let $\Gamma$ be now a connected graph with
$N+1$ vertices, H the corresponding Hamiltonian of the form
(\ref{Hdef}) and $|\psi>$ its ground state. By the Lemma, there
are two vertices each of which can be removed from $\Gamma$
without breaking the connectedness. Let us without loss of
generality renumber the vertices so that the removable ones will
be $1$ and $N+1$. Taking into account also (\ref{twospins}) we
see that (\ref{sisj}) holds for any $i,j=1,\dots,N$ and any
$i,j=2,\dots,N+1$ with $i\not= j$. So only ${\bf s_1\cdot
s_{N+1}}$ is under the question. As the total spin ${\bf S}^2$
commutes with the Hamiltonian $H$, we can separate $|\psi>$ into
the eigenstates of ${\bf S}^2$:
\begin{equation}
|\psi>=\sum_{S}|\psi_S>\;\;,\;\;
{\bf S}^2|\psi_S>=S(S+1)|\psi_S>\;\;,\;\; H|\psi_S>=0\;.
\end{equation}
Now
\begin{eqnarray}
\nonumber
{\bf S}^2|\psi_S> &=& \left[\left(\sum_{i=1}^N{\bf s_i}\right)^2
+ 2\sum_{i=1}^N{\bf s_i\cdot s_{N+1}} + {\bf s_{N+1}}^2\right]
|\psi_S>
\\ \nonumber
&=& \left[\frac{N}{2}\left(\frac{N}{2}+1\right) +
2{\bf s_1\cdot s_{N+1}} + 2\left(N-1\right)\frac{1}{4} +
\frac{3}{4}\right]|\psi_S>
\\
&=& \frac{1}{4}\left(N^2+4N+1\right)|\psi_S>
+ 2 {\bf s_1\cdot s_{N+1}}|\psi_S>\quad .
\end{eqnarray}
So $|\psi_S>$ is an eigenvector of ${\bf s_1\cdot s_{N+1}}$. The
possible eingenvalues are $\frac{1}{4}$ and $-\frac{3}{4}$, we
show that the second possibility leads to a contradiction. Let
us assume
\begin{equation}
\nonumber
{\bf s_1\cdot s_{N+1}}|\psi_S>=-\frac{3}{4}|\psi_S> \quad ,
\end{equation}
then
\begin{equation}
\label{ss34}
S(S+1)=\frac{1}{4}(N^2+4N-5)
\end{equation}
with the possible values
$S=\frac{N+1}{2},\frac{N-1}{2},\dots,\frac{1}{2}$ or $0$.

If $N=2n$ is even ($n\geq 1$) then $S=k+\frac{1}{2}$ with a
non-negative integer $k$. Substituting into (\ref{ss34}),
\begin{equation}
\label{Neven}
(n-k)(n+k+2)=2 \quad .
\end{equation}
Here the product of two integers equals $2$ so their values must
be $\pm 1$ and $\pm 2$. As the second integer is larger than $2$,
Equation (\ref{Neven}) cannot be satisfied.

If $N=2n+1$ is odd (integer $n\geq 1$) then $S=k$ is a
non-negative integer. Equation (\ref{ss34}) now reads
\begin{equation}
k(k+1)=n(n+3) \quad .
\end{equation}
Here necessarily $k>n$, so $k=n+m$, $m>0$ which gives
\begin{equation}
2(m-1)n+m(m+1)=0 \quad .
\end{equation}
However, $(m-1)n\geq 0$ and $m(m+1)>0$ so the equation cannot be
satisfied.

Therefore the only possibility is
${\bf s_1\cdot s_{N+1}}|\psi_S>=\frac{1}{4}|\psi_S>$
and then $S=\frac{N+1}{2}$ for $|\psi_S>\not= 0$. The statement
b) is now proved for $N+1$ spins and so it holds for any N by
induction. The Theorem is now completely proved.
\\ \hspace*{10.9cm} \mbox{\rule[-1.pt]{6pt}{10pt}}

\section{Concluding remark}
We presented here a rigorous proof of the generally accepted essential
uniqueness of the ferromagnetic ground state.
We assumed that the edges on which the ferromagnetic interaction
occurs form a connected graph $\Gamma$, no other assumption concerning
dimension or topology of the spin site lattice is needed. Typical
situation of nearest-neighbors interaction is included as we can
always draw a connected path through the all lattice sites formed
by the edges between nearest neighbors (except of peculiar cases
like two disconnected ferromagnets). On the coupling constants we
assume only $J_{ij}\geq 0$ for all $i,j$ and $J_{ij}>0$ for
$<i,j>\subset\Gamma$, so they need not be same for the all pairs
of spins, in particular, the case of interaction of nearest
neighbors of higher orders with weaker couplings is allowed.
There are no other assumptions then connectedness on the topology
of $\Gamma$, in particular, it is irrelevant whether the periodic
conditions are required or not for the cubic lattice or the linear
chain.
\\ \\
{\bf Acknowledgement}\\
The work is partly supported by grants A1048801 and ME099.

\end{document}